\documentclass[aoas,nameyear,dvips]{arximspdf}

\doi{10.1214/09-AOAS312F}
\referstodoi{10.1214/09-AOAS312}
\volume{3}
\issue{4}
\pubyear{2009}
\firstpage{1295}
\lastpage{1298}

\begin{document}
\begin{frontmatter}

\title{Discussion of: Brownian distance covariance}
\runtitle{Discussion}
\pdftitle{Discussion on Brownian distance covariance by G. J. Szekely and M. L. Rizzo}

\begin{aug}
\author[A]{\fnms{Bruno} \snm{R\'{e}millard}\corref{}\ead[label=e1]{bruno.remillard@hec.ca}}
\runauthor{B. R\'{e}millard}
\affiliation{HEC Montr\'{e}al}
\address[A]{GERAD and Service de l'enseignement
\\des m\'{e}thodes quantitatives de gestion \\ HEC Montr\'{e}al \\
Montr\'{e}al\\ Canada H3T 2A7} 
\end{aug}




\end{frontmatter}

In \citet{SzekelyRizzoBakirov2007}, the notions of distance
covariance and distance correlation between two random vectors were
introduced. It was shown that the distance covariance is zero if and
only if the two vectors were independent. An empirical version was
also defined and its limiting distribution was investigated, under
the null hypothesis of independence; furthermore, the underlying
test based on the empirical version of the distance covariance is
consistent in the sense that under the hypothesis of dependence, its
power tends to one as the sample size tends to infinity.

In the present paper the authors continue the study of the
properties of the distance covariance and they show that it can be
defined in terms of covariances of multivariate Brownian processes.
They also generalized that idea to other stochastic processes,
namely, multivariate fractional Brownian motions. Defining dependence
measures through other stochastic processes is quite interesting, but
except for the few cases stated in the paper, it is still to be proven
useful. I encourage the authors to continue to explore that
interesting idea. Here are some questions I would like to be answered:
(i) Can other dependence measures be written in that form, for example,
Kendall's
tau? (ii) What are the conditions on the underlying processes so that
the value of the covariance is zero if and only if the two random
vectors are independent? (iii) Can you prove a central limit theorem
for the empirical version and what are the conditions on the underlying
stochastic processes for the existence of the limiting distribution?

In what follows I will suggest some other extensions and applications
of the notion of covariance distance and distance correlation. More
precisely, I will describe extensions using rank-based methods and
suggest two applications in a multivariate context, that is, when more
than two random vectors are involved.

\section{Rank-based methods}
In my opinion, there are two weaknesses of the distance covariance: The
moment assumption on the random vectors and the fact that the
dependence measure depends on the marginal distributions. That problem
can be dealt with easily when the margins are continuous by using the
associated uniform variables defined through the well-known mapping
\[
X^{(j)} \mapsto U^{(j)} = F_{X^{(j)}}\bigl(X^{(j)}\bigr), \qquad  j=1,\ldots, p,
\]
\[
Y^{(k)} \mapsto V^{(k)} = F_{Y^{(k)}}\bigl(Y^{(k)}\bigr), \qquad  k=1,\ldots,
q.
\]

\noindent
Then, the distance covariance between $U=(U^{(1)},\ldots, U^{(p)})$ and
$V = (V^{(1)},  \ldots,\break V^{(q)})$ only depends on the underlying
copula of $(U,V)$ and $X$ and $Y$ are independent if and only if $U$
and $V$ are independent. Its empirical counterpart is simply computed
by replacing the observations by their normalized ranks, that is,
replacing $(X_i,Y_i)$ by $ (R_{X,i}/n, R_{Y,i}/n)$, where
$R_{X,ij}$ is the rank of $X_{ij}$ among $X_{1j}, \ldots, X_{nj}$,
$j=1,\ldots, p$. It is relatively easy to prove that the limiting
distribution of
$n\mathcal{V}_n^2 (U,V)$ will converge to $\|\xi\|^2$, where the
covariance of $\xi$ is $R_{U,V}$, as has been defined in Theorem 5.

On the subject of rank-based methods, I disagree with the authors when
they say that these methods are effective only for testing linear or
monotone types of dependence. Because
independence can also be characterized by copulas, and the latter can
be efficiently estimated with ranks, their statement is totally
inadequate. See, for example, \citet{GenestRemillard2004} for
tests of nonserial and serial dependance based on ranks. Furthermore,
in their Example 2, the authors suggest that the test based on the
distance covariance is more powerful that its rank-based analog.
Looking at Figure 2, this is the case only when the sample size $n$ is
quite small ($ n \le15$). I would be more convinced by a simulation
with different dependence models and sample sizes of the order 50 or
100, at the very least.

\section{Measuring dependence between several random vectors}
As a competitor to the distance covariance for tests of independence,
it is worth mentioning the Cram\'{e}r--von Mises statistic $n B_n$, where
\[
B_n = \int_{\mathbb{R}^{p+q}}\{F_{X,Y}^n (x,y)-F_{X}^n (x)F_{Y}^n (y) \}
^2 \,d F_{X,Y}^n(x,y)
\]
is the empirical counterpart of
\[
B = \int_{\mathbb{R}^{p+q}}\{ F_{X,Y}(x,y)-F_X(x)F_Y(y) \}^2\, d F_{X,Y}(x,y).
\]
The latter dependence measure also characterizes independence in the
sense that $B=0$ only when $X$ and $Y$ are independent.

The limiting distribution of $ n^{1/2}\{ F_{X,Y}^n(x,y)-
F_{X}^n(x)F_{Y}^n(y)\}$ used to construct $B_n$ was studied in
\citet{BeranBilodeauLafaye2007}. In fact, the authors proposed
testing independence between $d$ random vectors $Z_1, \ldots, Z_d$,
using statistics based on $ \mathbb{F}_n = n^{1/2}\{ H_n(z_1,\ldots,
z_d)-\break F_{n,1}(z_1) \cdots F_{n,d}(z_d)\}$, where $H_n$ is the
empirical joint distribution function of $(Z_1, \ldots, Z_d)$, and
$F_{n,j}$ is the empirical joint distribution of $Z_j$, $j\in
\{1,\ldots, d\}$, calculated from a sample $(Z_{11}, \ldots,
Z_{1d}), \ldots, (Z_{n1}, \ldots, Z_{nd})$. Extending the results of
\citet{GhoudiKulpergerRemillard2001} from random variables to
random vectors, \citet{BeranBilodeauLafaye2007} considered tests
of nonserial and serial dependence based on M\"{o}bius decomposition
of $\mathbb{F}_n$, yielding asymptotically independent empirical
processes $\mathbb{F}_{n,A}$ (depending only on the indices in $A$),
for any subset $A$ of $\{1, \ldots, d\}$ containing at least two
elements. These $2^d-d-1$ processes can be combined to define
powerful tests of independence [\citet{GenestQuessyRemillard2007}].

Because the limiting distribution under the null hypothesis depends
on the unknown distribution function $F_1,\ldots, F_d$,
Beran, Bilodeau and Lafaye~de
Micheaux (\citeyear{BeranBilodeauLafaye2007}) showed that bootstrap methods
worked for estimating the $P$-value of underlying test statistics.

Further, note that \citet{BilodeauLafaye2005} defined tests on
independence between random vectors based on characteristic
functions, when the marginal distributions were assumed to be
Gaussian. They considered both serial and nonserial cases. The
Cram\'{e}r--von Mises type statistics they used are quite similar to
the statistic $n\mathcal{V}_n^2$, when restricted to two random
vectors. Therefore, it would be worth considering distance covariance
measures for measuring independence between several random vectors.
In order to get nice covariance structures, M\"{o}bius
transformations of the empirical characteristic functions should be
used. More precisely, for any $A \subset\{1,\ldots, d\}$, one could
define distance covariance measures $\mathcal{V}_{n,A} =
\|\xi_{n,A}\|^2$, where
\[
\xi_{n,A}(t_1,\ldots, t_d) = n^{-1/2}\sum_{j=1}^n \prod_{k \in A}
\bigl\{e^{i\langle t_k,Z_{jk}\rangle}-f_{X_k}^n (t_k)\bigr\}.
\]

\section{Measuring dependence for multivariate time series}

The distance covariance measures should also be defined in a time
series context to measure serial dependence. For example, if
$(Z_i)_{i\ge1}$ is a stationary multivariate time series, one can
easily define the ``distance autocovariance'' by
\[
\mathcal{V}^2(l) = \mathcal{V}^2(Z_j, Z_{j+l}), \qquad l \ge1.
\]

It is easy to show that under the white noise hypothesis and the
assumption that $|Z_1|_p$ has finite expectation,
\[
n \mathcal{V}_n^2(l) \stackrel{D}{\longrightarrow} \|\xi_l\|^2,
\]
where $\xi_1, \ldots, \xi_m$ are independent copies of $\xi$, as
defined in Theorem 5. Again, M\"{o}bius transformations should be used
to test independence between $(Z_1, \ldots, Z_m)$. Therefore, there
are still many interesting avenues to explore, especially for time
series applications. For example, rank-based methods could also be
used. See, for example, \citet{GenestRemillard2004}.

\section{Using residuals and pseudo-observations}
Finally, one could ask what happens when observations are replaced
by residuals (or pseudo-observations like normalized ranks)? For
example, one would like to test independence of the error terms in
several linear models, using the residuals. Based on the results in
\citet{GhoudiKulpergerRemillard2001}, the limiting distribution
of $n\mathcal{V}_n^2$ should remain the same, under weak
assumptions. That should also be true for the multidimensional
extensions of the distance covariance. However, replacing the
unobservable innovations by residuals in multivariate time series
models leads to completely different limiting processes. For
example, using residuals of a simple AR(1) model of the form $Z_t =
\mu+\phi(Z_{t-1}-\mu)+\varepsilon_t$, one can show that
$n\mathcal{V}_n^2(l)$ converges in law to $\|\xi_l-\gamma_l\|^2$,
where
\[
\gamma_l(t,s) = sf(s)f'(t)\Phi\phi^{l-1},
\]
where $f$ is the characteristic function of $\varepsilon_t$, and
$\phi_n$ is an estimator of $\phi$ so that $n^{1/2}(\phi_n-\phi)$
converges in law to $\Phi$.

Fortunately, using an analog of the transform $\Psi$ defined in\break
\citeauthor{GenestGhoudiRemillard2007} [(\citeyear{GenestGhoudiRemillard2007}), page 1373], it might be possible
to obtain limiting distributions not depending on the estimated
parameters.

%



%

\printaddresses


\begin{thebibliography}{}

\bibitem[\protect\citeauthoryear{Beran, Bilodeau and Lafaye~de
Micheaux}{2007}]{BeranBilodeauLafaye2007}
\textsc{Beran, R., Bilodeau, M.} and \textsc{Lafaye de Micheaux,  P.} (2007).
Nonparametric tests of independence between random vectors.
\textit{J. Multivariate Anal.} \textbf{98} 1805--1824.
\MR{2392434}

\bibitem[\protect\citeauthoryear{Bilodeau and Lafaye de
Micheaux}{2005}]{BilodeauLafaye2005}
\textsc{Bilodeau, M.} and \textsc{Lafaye de Micheaux,  P.} (2005).
A multivariate empirical characteristic function test of independence
with normal marginals.
\textit{J. Multivariate Anal.} \textbf{95} 345--369.
\MR{2170401}

\bibitem[\protect\citeauthoryear{Genest, Ghoudi and R\'
{e}millard}{2007}]{GenestGhoudiRemillard2007}
\textsc{Genest, C., Ghoudi, K.} and \textsc{R\'{e}millard, B.} (2007).
Rank-based extensions of the {B}rock {D}echert {S}cheinkman test for
serial dependence.
\textit{J. Amer. Statist. Assoc.} \textbf{102} 1363--1376.
\MR{2372539}

\bibitem[\protect\citeauthoryear{Genest, Quessy and R\'{e}millard}{2007}]{GenestQuessyRemillard2007}
\textsc{Genest, C., Quessy, J.-F.} and \textsc{R\'{e}millard, B.} (2007).
Asymptotic local efficiency of {C}ram\'{e}r--von {M}ises tests for
multivariate independence.
\textit{Ann. Statist.} \textbf{35} 166--191.
\MR{2332273}

\bibitem[\protect\citeauthoryear{Genest and R\'
{e}millard}{2004}]{GenestRemillard2004}
\textsc{Genest, C.} and \textsc{R\'{e}millard, B.} (2004).
Tests of independence or randomness based on the empirical copula
process. \textit{Test} \textbf{13} 335--370.
\MR{2154005}

\bibitem[\protect\citeauthoryear{Ghoudi, Kulperger and R\'
{e}millard}{2001}]{GhoudiKulpergerRemillard2001}
\textsc{Ghoudi, K., Kulperger, R. J.} and \textsc{R\'{e}millard, B.} (2001).
A nonparametric test of serial independence for time series and
residuals.
\textit{J. Multivariate Anal.} \textbf{79} 191--218.
\MR{1868288}

\bibitem[\protect\citeauthoryear{Sz{\'{e}}kely, Rizzo and
Bakirov}{2007}]{SzekelyRizzoBakirov2007}
\textsc{Sz{\'{e}}kely, G. J., Rizzo, M. L.} and \textsc{Bakirov, N. K.} (2007).
Measuring and testing dependence by correlation of distances.
\textit{Ann. Statist.} \textbf{35} 2769--2794.
\MR{2382665}

\end{thebibliography}
\end{document}